# Energy transfer up-conversion in $Tm^{3+}$- doped silica fibre

D. A. Simpson, G. W. Baxter and S. F. Collins

*Centre for Telecommunications and Micro-Electronics, Optical Technology Research Laboratory, Victoria University, PO Box 14428 Melbourne, VIC 8001, Australia.*

W. E. K. Gibbs

*Centre for Imaging and Applied Optics, Swinburne University, VIC 3122, Australia.*

W. Blanc, B. Dussardier and G. Monnom

*Laboratoire de Physique de la Matière Condensée UMR6622, Université de Nice - Sophia Antipolis Parc Valrose, 06108 Nice, Cedex 2, France.*

**Abstract**

A study of the mechanisms responsible for the infra-red to near infra-red up-conversion in thulium doped silica fibres is presented. Up-conversion luminescence was observed from the $^3H_4$ level of $Tm^{3+}$ under 1586 nm pumping into the $^3F_4$ level. The quadratic dependence of the up-conversion luminescence at 800 nm on the 1800 nm luminescence from the $^3F_4$ level confirms that the $^3H_4$ level is populated by a two photon process. Two possible processes are proposed as mechanisms responsible for the up-conversion: excited state absorption and energy transfer up-conversion. The decay characteristics of the luminescence from the $^3H_4$ level were studied under direct and indirect pumping at 786 and 1586 nm, respectively. By comparing the decay waveforms to the solution of a simple set of rate equations, the energy transfer up-conversion process ($^3F_4, ^3F_4 \rightarrow ^3H_4, ^3H_6$) is established at $Tm_2O_3$ concentrations greater than 35 ppm mol.

1. Introduction

The broadband emission from the $^3H_4 \rightarrow {}^3F_4$ transition in $Tm^{3+}$ has been identified as one of the more promising candidates for optical amplification in the telecommunication S-Band (1470 - 1530 nm). Amplification in the S-band region has been observed in many $Tm^{3+}$ doped fluoride crystals [1-4]. The low phonon energies associated with fluoride crystals allow many of the thulium transitions to operate with quantum efficiencies near 100%. When doped into glasses with higher phonon energies such as silica, the quantum efficiencies of these transitions are reduced, in some cases, to a few percent. Although recent work in thulium doped silica fibres showed a four fold increase in the quantum efficiency of the $^3H_4 \rightarrow {}^3F_4$ amplifying transition through the incorporation of aluminium, this is still considerably less than the almost 100% efficiency observed in $Tm^{3+}$ doped fluoride glasses. If further improvements in the efficiency of the thulium transitions are to be realised in silica glass, an understanding of the spectroscopic processes involved in the related levels is required.

Extensive studies have been carried out on the cross relaxation process ($^3H_4,{}^3H_6 \rightarrow {}^3F_4,{}^3F_4$) originating from the $^3H_4$ level in various thulium doped crystals and oxide glasses [5-7]. However, little attention has been given to the energy transfer up-conversion process originating from the $^3F_4$ level ($^3F_4,{}^3F_4 \rightarrow {}^3H_4,{}^3H_6$). This process has been observed in thulium doped crystals [8-10]. However, in some cases, the sharp spectral lines associated with these materials results in a lack of spectral overlap between the $^3F_4 \rightarrow {}^3H_4$ absorption and the $^3F_4 \rightarrow {}^3H_6$ emission cross section, which reduces the efficiency of the energy transfer process. In silica glasses, where the energy levels are inhomogeneously broadened, the spectral overlap between the absorption and emission cross sections is expected to be much stronger and therefore the effect of energy transfer up-conversion is expected to be much greater. Work done in heavily thulium doped silica fibres by Jackson has attributed quenching of the $^3F_4$ level's population to the ($^3F_4,{}^3F_4 \rightarrow {}^3H_4,{}^3H_6$) energy transfer process and another possible process involving the $^3H_5$ level ($^3F_4,^3F_4 \rightarrow ^3H_5,^3H_6$) [11]. Here the energy transfer up-conversion process ($^3F_4,^3F_4 \rightarrow ^3H_4,^3H_6$) is established and studied in thulium doped silica fibres at low $Tm_2O_3$ concentrations.

Room temperature up-conversion luminescence has been observed from the $^3H_4$ level of thulium under 1586 nm pumping into the $^3F_4$ level. Spectroscopic methods were used to investigate the mechanisms responsible for this up-conversion in fibre samples with three different concentrations (35, 105 and 515 ppm mol $Tm_2O_3$).

## 2. Experimental Details

The samples used in this investigation were fabricated using the modified chemical vapour deposition and solution doping techniques. The three alumino-silicate samples had $Tm_2O_3$ concentrations of 35, 105 and 515 ppm mol and contained 4.8, 4.5 and 4.6 mol% of $Al_2O_3$, respectively. The $Tm_2O_3$ concentration was estimated by measuring the absorption peak at 780 nm and using a cross-section value of ($8.7 \times 10^{-25}$ m$^2$) [12]. The concentration in at/m$^3$ was converted into ppm mol assuming that the silica density is 2.2 [13]. A York S14 index profiler was used to obtain the refractive index profiles of the fibre samples and the $Al_2O_3$ concentrations were estimated from these profiles since $Al_2O_3$ has been shown to increase the optical index of silica by $2.3 \times 10^{-3}$ per mol % [14]. It was assumed that the concentration of $Tm^{3+}$ did not contribute significantly to the index difference. No other standard modifiers of silica, such as Ge, P, F, were used in the fabrication of these samples.

A pigtailed Alcatel laser diode provided excitation at 1586 nm to study the steady state luminescence intensities. An Avtech laser driver ran the diode from 1-15 mW in quasi-cw mode with 12.5 ms pulses at a repetition rate of 40 Hz to allow for lock-in detection. As the 12.5 ms pulse length is at least 10 times longer than the longest rise time of any excited energy level in the thulium-doped silica system, the intensity of the luminescence at 800 and 1800 nm represents the intensity under steady state conditions.

The luminescence was collected immediately after the splice to the excitation source with a collimating lens transverse to the doped fibre. Sample lengths were kept to 50 mm to minimise the effects of amplified spontaneous emission and re-absorption. The up-conversion luminescence at 800 nm was directed through a 900 nm short pass filter and detected using a Hamamatsu R928 photomultiplier. The luminescence at 1800 nm was directed through a 1500 nm long pass filter and detected using a Thorlabs (FG20) InGaAs photodiode.

Decay measurements were carried out using the same arrangement as described above, but with the lock-in amplifier replaced by a Tektronix TDS320 digital oscilloscope. Luminescence decay waveforms were recorded under 786 and 1586 nm pump excitation. The 12.5 ms pulses at a repetition rate of 40 Hz at 786 nm were provided by a Spectra Physics argon-ion pumped Ti-sapphire laser in conjunction with an acousto-optic modulator. The 800 nm and 1800 nm detection systems had response times of 0.8 and 15 μs, respectively. The luminescence decay waveforms were averaged 256 times before being sent to a PC for processing.

### 3. Results:

The counter propagating up-conversion luminescence spectrum from the $^3H_4$ level of thulium when excited at 1586 nm is shown in Figure 1. The emission spectrum observed is characteristic of the $^3H_4$ level. The large dip in the spectrum at ~ 795 nm, is assigned to the strong ground state absorption ($^3H_6 \rightarrow {}^3H_4$).

The origin of this luminescence lies in two possible processes, excited state absorption (ESA) and energy transfer up-conversion (ETU). In ESA, a pump photon excites an ion to the $^3F_4$ level. The excited ion then absorbs the energy of another pump photon promoting itself to the $^3H_4$ level. In ETU, two pump photons excite two ions to the $^3F_4$ level. The excited ions then exchange energy non-radiatively, promoting one excited ion to the $^3H_4$ level, whilst demoting the other ion to the $^3H_6$ ground state (see Figure 2). It should be noted that there is an energy mismatch between the $^3F_4 \rightarrow {}^3H_4$ absorption and the $^3F_4 \rightarrow {}^3H_6$ emission that necessitates the assistance of phonons. The minimum energy gap required to complete the transition in Tm$^{3+}$:YCL$_3$ was -698 cm$^{-1}$ (the negative sign indicating an endothermic process) [15], in silica glass this energy gap can be bridged with the absorption of just one vibrational mode of the glass.

Both processes illustrated in Figure 2 require two pump photons in the process of exciting one ion to the $^3H_4$ level. Hence, the luminescence from the $^3H_4$ level at 800 nm should be proportional to the square of the population in the $^3F_4$ level. i.e. proportional to the square of the luminescence from this level at 1800 nm. This was investigated by measuring the steady state luminescence intensities at 800 and 1800 nm over a range of pump powers. Figure 3 shows the dependence of the 800 nm luminescence on the 1800 nm luminescence for the three samples over the available range of pump powers. The approximate quadratic dependence observed for all samples is consistent with the $^3H_4$ level being populated by a two photon process.

*3.1 Direct pumping of the $^3H_4$ level*

Luminescence decay waveforms from the $^3H_4$ level were recorded under direct excitation at 786 nm with an incident pump power of 1.6 mW. The luminescence decay waveforms were non-exponential; therefore the lifetime of the $^3H_4$ level was taken as the time required for the fluorescence intensity to decrease to 1/*e* of its peak intensity. This non-exponential behaviour has been observed in many other thulium doped glasses [5, 16, 17]. In many cases, the non exponential nature of the decay is a result of energy transfer processes and diffusion occurring from the $^3H_4$ level [18]. However, Lincoln states that in the absence of energy transfer processes this non-exponential behaviour can be attributed to the inhomogeneous broadening of the atomic levels in silica [19]. The 1/*e* lifetime for each sample including the fitting errors is listed in Table 1.

The measured lifetime of the $^3H_4$ level in these thulium samples is considerably less than the radiative lifetime expected in silica glass (~ 650 µs) [20], due to the dominance of multi-phonon decay from the $^3H_4$ level to the $^3H_5$ (then the $^3F_4$ level). This multi-phonon decay limits the radiative quantum efficiency of the $^3H_4$ level to ~ 5 %.

It is noted that the decay constant of the $^3H_4$ level does not appear to be affected by the well known cross relaxation process ($^3H_4,^3H_6 \rightarrow ^3F_4,^3F_4$). If this cross relaxation process was operating efficiently in these samples a decrease in the $^3H_4$ level's decay constant with increasing concentration would be evident. Recent work on these types of silica fibres has shown that the decay constant of the $^3H_4$ level is more sensitive to the amount of aluminium incorporated into the fibre's core [21].

*3.2 Indirect pumping of the $^3H_4$ level*

Indirect pumping of the $^3H_4$ level was achieved by exciting the fibre samples at 1586 nm. The up-conversion luminescence decay characteristics were studied to ascertain the mechanism/s responsible for populating the $^3H_4$ level. Figure 4 shows the semi-log plot of the up-conversion luminescence decay under 1586 nm pumping for each sample and, as a comparison, the decay of the $^3H_4$ level under direct excitation is presented also.

The luminescence decay from the $^3F_4$ level was also studied under 1586 nm pumping; the decay from that level at 1800 nm was characterised by a single exponential. The measured $^3F_4$ lifetime for each sample is listed in table 1.

## 4. Discussion:

The behaviour of the up-conversion luminescence when the pump excitation has been removed can be described by considering a simple set of rate equations relating the populations of the $^3F_4$ and $^3H_4$ energy levels, namely

$$\frac{dn_1}{dt} = -\frac{n_1}{\tau_1} - W_{ETU} n_1^2 c \qquad (1)$$

$$\frac{dn_2}{dt} = -\frac{n_2}{\tau_2} + \frac{1}{2} W_{ETU} n_1^2 c, \qquad (2)$$

where $n_1$ and $n_2$ represent the normalised population of the $^3F_4$ and $^3H_4$ levels respectively, $c$ is the $Tm^{3+}$ concentration, $W_{ETU}$ is the energy transfer up-conversion co-efficient and $\tau_1$ and $\tau_2$ are the lifetimes of the $^3F_4$ and $^3H_4$ levels, respectively.

These rate equations involve several assumptions. Firstly, it has been assumed that any population in the $^3H_5$ level will relax rapidly to the $^3F_4$ level in a time scale which is short in comparison to the other decay times involved; thus the presence of the $^3H_5$ level has been ignored. Secondly, by representing the ETU process with the term $W_{ETU} n_1^2$ we have ignored any energy migration processes, which is justifiable since this process occurs on a much smaller time scale (~$10^{-10}$ s) [22]. Thirdly, it was assumed that ESA of the pump and signal photons can be ignored due to the relatively low ESA cross sections at the respective wavelengths. Peterka *et al* [23], have estimated the ESA cross section at the pump and signal wavelengths (1586 and 1800 nm) to be ~ $3\times10^{-31}$ and ~ 0 m², respectively. Although there is a relatively large error associated with these values, it suggests that ESA does not play a significant role in the up-conversion process at these wavelengths.

It has been noted in section 3.2 that the luminescence decay from the $^3F_4$ level was a single exponential, implying $\frac{n_1}{\tau_1} \gg W_{ETU} n_1^2$. Therefore the solution to equation (1) becomes:

$$n_1 = n_{10} \exp\left(-\frac{t}{\tau_1}\right), \tag{3}$$

where $n_{10}$ represents the initial population of the $^3F_4$ level after the pump excitation is removed.

By inserting equation (3) into (2) we can solve for $n_2$, giving

$$n_2 = \left(n_{20} - \frac{A}{\tau_2^{-1} - 2\tau_1^{-1}}\right)\exp\left(-\frac{t}{\tau_2}\right) + \frac{A}{\tau_2^{-1} - 2\tau_1^{-1}}\exp\left(-\frac{2t}{\tau_1}\right), \tag{4}$$

where $A = \frac{W_{ETU} n_{10}^2 c}{2}$, and $n_{20}$ represents the initial population of the $^3H_4$ level after the pump excitation is removed.

From Equation (4) we see that the up-conversion luminescence, resulting from ETU, decays as a double exponential with two characteristic time constants: one being equal to the lifetime of the $^3H_4$ level ($\tau_2$) and the other being equal to ($\tau_1/2$) i.e. one half of the $^3F_4$ level lifetime.

It is clear in Figure 4 that the up-conversion luminescence decay has two exponential components. A least squares fit to Equation (4) was applied to the up-conversion luminescence decay waveforms, with $\tau_2$ being the 1/e lifetime for the $^3H_4$ level (obtained under direct pumping), leaving $\tau_1$ and $A$ as the fitting parameters. The least squares fit was in excellent agreement with the measured waveforms. Table 1 shows that the fitted and measured values of $\tau_1$ are also in good agreement verifying the validity of this analysis.

The ETU exhibited in these samples suggests the presence of ion pairs or clusters. These results imply that, even at low $Tm_2O_3$ concentrations, ion pairs can still be formed in alumino-silicate glasses. The absence of a glass softening agent such as phosphorus may have also contributed to the forming of ion pairs. The spectroscopic measurements reported here do not allow the determination of the relative strength of the ETU process; this measurement is planned.

## 5. Conclusion

Up-conversion luminescence from the $^3H_4$ level in thulium-doped silica fibres, when excited at 1586 nm has been identified. The up-conversion luminescence from the $^3H_4 \rightarrow {}^3H_6$ transition showed a quadratic dependence on the luminescence from the $^3F_4 \rightarrow {}^3H_6$ transition, indicating a two photon up-conversion process. The double exponential decay of the up-conversion luminescence confirmed that ETU was operating at $Tm_2O_3$ concentrations as low as 35 ppm mol. The fitted value of $\tau_1$ obtained from the rate equation analysis is in good agreement with the measured $\tau_1$ value for all samples studied. The presence of ETU at $Tm_2O_3$ concentrations greater than 35 ppm mol will be an important consideration when designing and fabricating active thulium-doped silica fibre devices in the future.

**Acknowledgements**

The authors gratefully acknowledge helpful discussions with Dr. Pavel Peterka from the Institute of Radio Engineering and Electronics in Prague, Czech Republic. This work was supported by the Australian Research Council, and Centre National de la Recherche Scientifique, in France.

**References**

[1] S. Tanabe, T. Tamaoka, J. Non-Cryst. Solids 326-327 (2003) 283
[2] M. Shi Qing, W. Sun Fat, E. Y. B. Pun, C. Po Shuen, J. Opt. Soc. Am. B, Opt. Phys. 21 (2004) 313
[3] L. Dong Jun, H. Jong, P. Se Ho, J. Non-Cryst. Solids 331 (2003) 184
[4] A. M. Jurdyc, G. Rault, W. Meffre, J. Le Person, S. Guy, F. Smektala, J. L. Adam, Proc. SPIE - Int. Soc. Opt. Eng. 4645 (2002) 79
[5] A. S. S. de Camargo, S. L. de Oliveira, D. F. de Sousa, L. A. O. Nunes, D. W. Hewak, J. Phys., Condens. Matter. 14 (2002) 9495
[6] I. R. Martin, V. D. Rodriguez, R. Alcala, R. Cases, J. Non-Cryst. Solids 161 (1993) 294
[7] A. Sennaroglu, A. Kurt, G. Ozen, J. Phys., Condens. Matter. 16 (2004) 2471

[8]  J. B. Gruber, M. E. Hills, R. M. Macfarlane, C. A. Morrison, G. A. Turner, G. J. Quarles, G. J. Kintz, L. Esterowitz, Phys. Rev. B, Condens. Matter 40 (1989) 9464
[9]  J. B. Gruber, M. D. Seltzer, M. E. Hills, S. B. Stevens, C. A. Morrison, J. Appl. Phys. 73 (1993) 1929
[10] L. B. Shaw, R. S. F. Chang, N. Djeu, Phys. Rev. B, Condens. Matter 50 (1994) 6609
[11] S. D. Jackson, Opt. Commun. 230 (2004) 197
[12] D. C. Hanna, I. R. Perry, J. R. Lincoln, J. E. Townsend, Opt. Commun. 80 (1990) 52
[13] H. Kakiuchida, N. Shimodaira, E. H. Sekiya, K. Saito, A. J. Ikushima, Appl. Phys. Lett. 86 (2005) 161907
[14] G. G. Vienne, W. S. Brocklesby, R. S. Brown, Z. J. Chen, J. D. Minelly, J. E. Roman, D. N. Payne, Opt. Fiber Technol., Mater. Devices Syst. 2 (1996) 387
[15] J. Ganem, J. Crawford, P. Schmidt, N. W. Jenkins, S. R. Bowman, Phys. Rev., B, Condens, Matter Mater. Phys. 66 (2002) 245101
[16] M. Bettinelli, F. S. Ermeneux, R. Moncorge, E. Cavalli, J. Phys., Condens. Matter. 10 (1998) 8207
[17] J. R. Lincoln, W. S. Brocklesby, F. Cusso, J. E. Townsend, A. C. Tropper, A. Pearson, J. Lumin. 50 (1991) 297
[18] A. Brenier, C. Pedrini, B. Moine, J. L. Adam, C. Pledel, Phys. Rev. B, Condens. Matter 41 (1990) 5364
[19] J. Lincoln, University of Southampton, (1992)
[20] M. J. F. Digonnet, Rare earth doped fiber lasers and amplifiers, 2nd Ed., Marcel Dekker, Inc., New York, (2001)
[21] B. Faure, W. Blanc, B. Dussardier, G. Monnom, P. Peterka, Technical Digest of Optical Amplifiers and Their Applications, San Francisco, (2004), OWC2
[22] D. L. Drexter, J. H. Schulman, 22 (1964) 1063
[23] P. Peterka, B. Faure, W. Blanc, M. Karasek, B. Dussardier, Opt. Quantum Electron. 36 (2004) 201

List of Tables:

Table 1: Measured and fitted lifetimes of the $^3H_4$ and $^3F_4$ energy levels under direct and indirect pumping.

| $Tm_2O_3$ conc.(ppm mol) | | 35 | 105 | 515 |
|---|---|---|---|---|
| $^3H_4$ lifetime (µs) $\lambda_p$= 786 nm | ($\tau_2$) (measured) | 28.3 ± 0.7 | 28.2 ± 0.6 | 31.0 ± 0.5 |
| $^3F_4$ lifetime (µs) $\lambda_p$= 1586 nm | ($\tau_1$) (fitted) | 482 ± 8 | 466 ± 2 | 460 ± 2 |
| | ($\tau_1$) (measured) | 475 ± 3 | 459 ± 2 | 473 ± 2 |

List of figures:

**Figure 1**

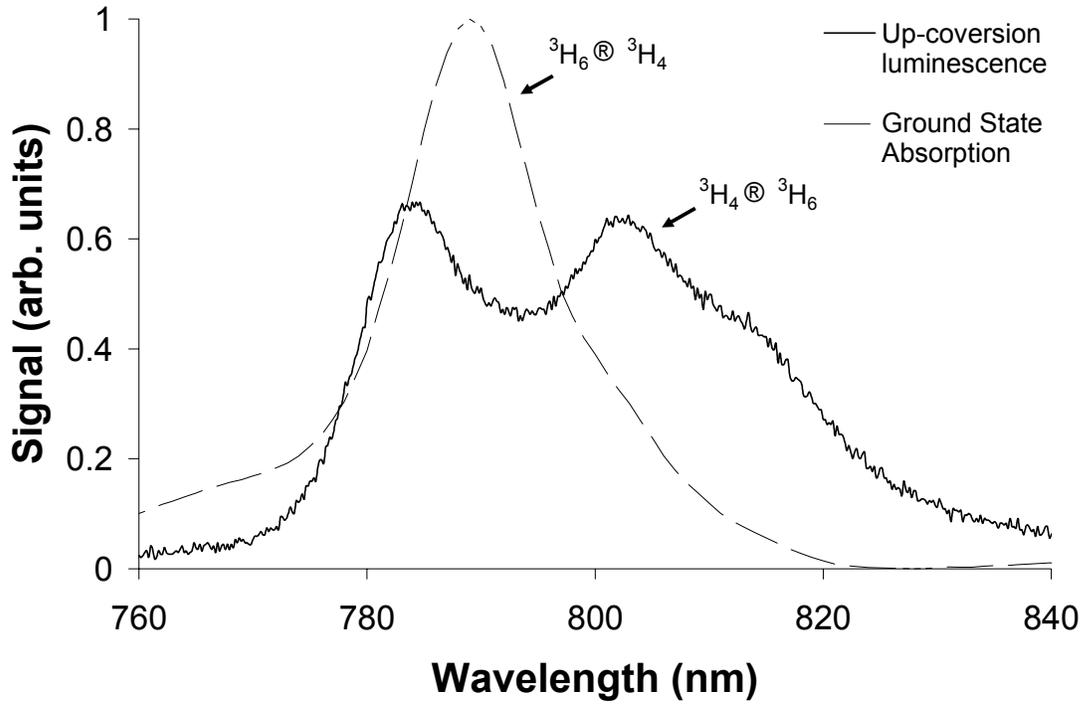

**Figure 2**

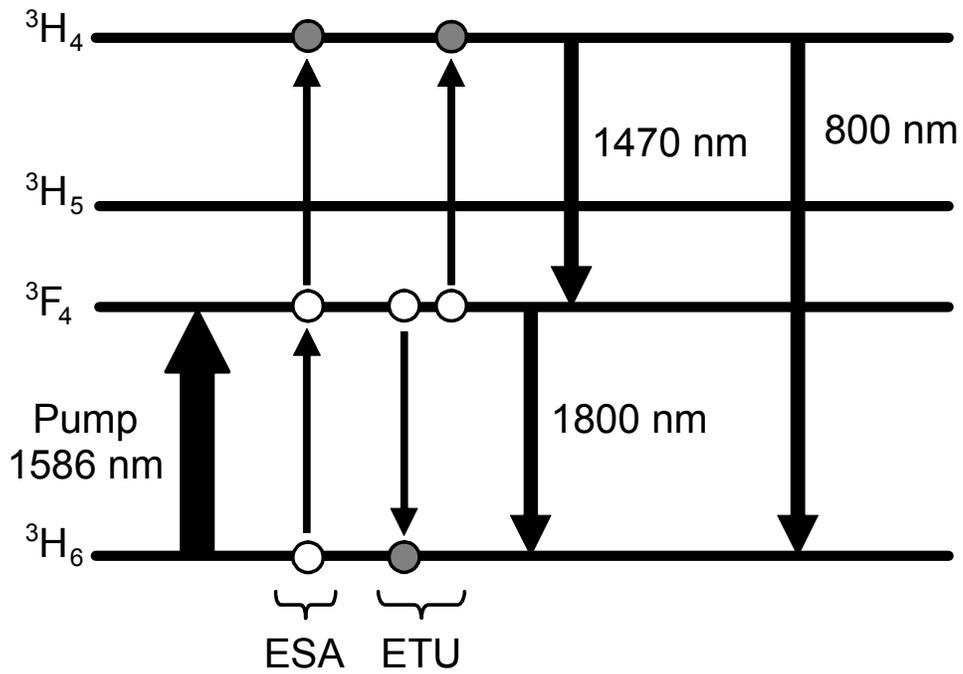

**Figure 3**

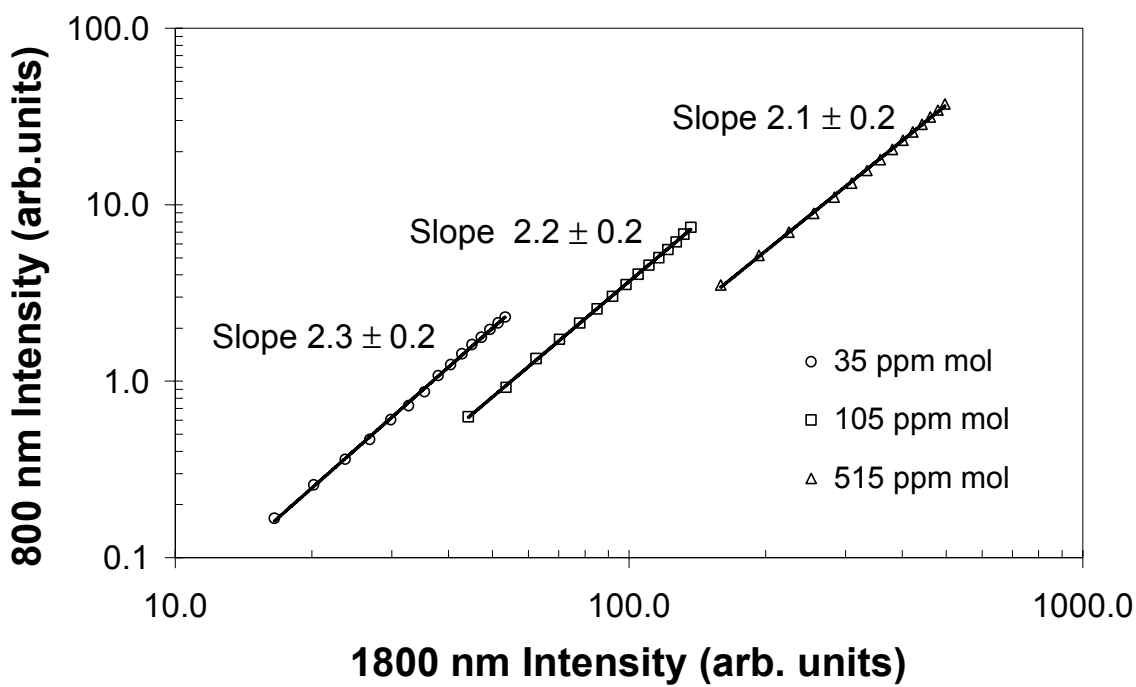

**Figure 4**

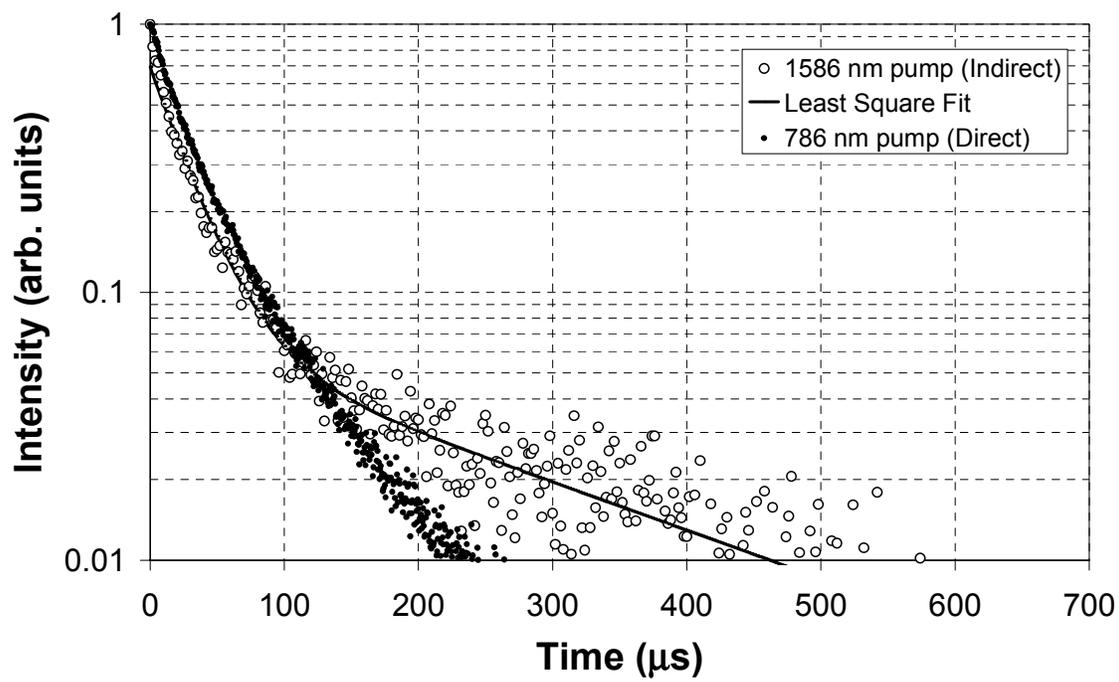

(a)

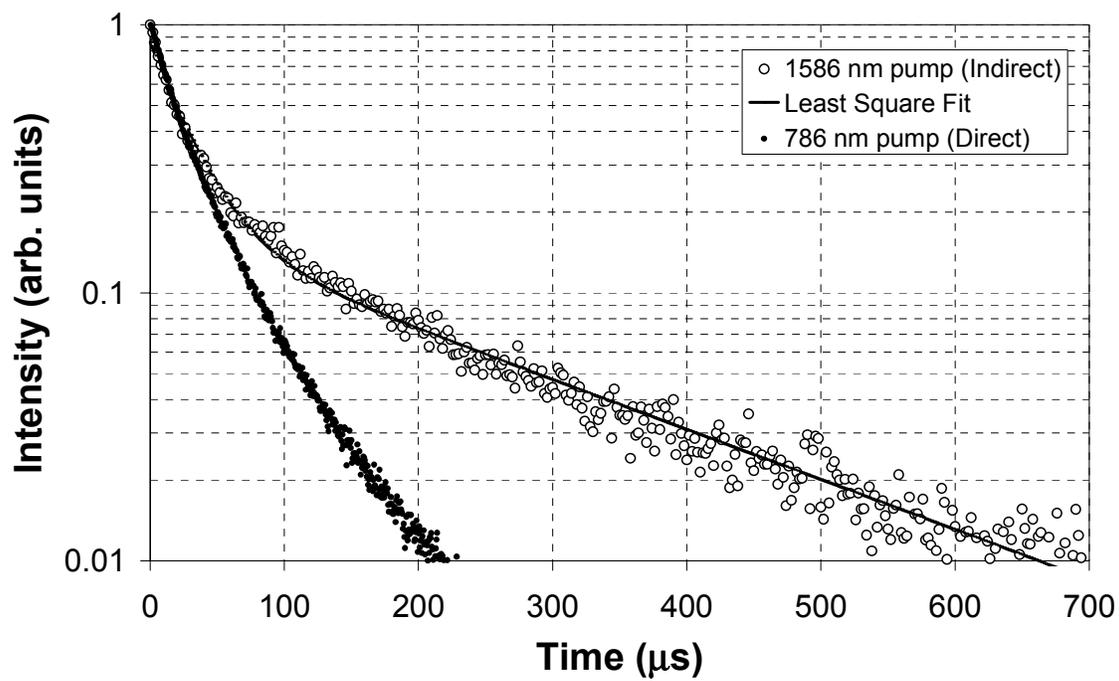

(b)

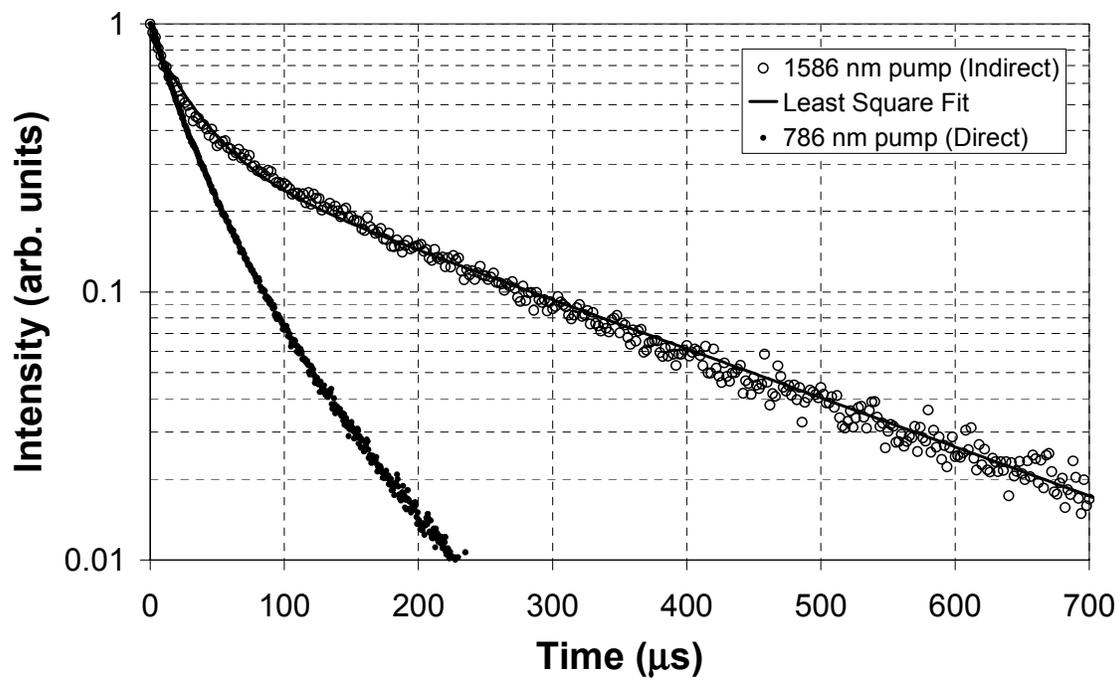

(c)

Figure Captions

Figure 1. Counter-propagating up-conversion luminescence spectrum from a 20 cm section of thulium doped silica fibre excited at 1586 nm.

Figure 2. Possible energy transfer processes that can populate the $^3H_4$ level in thulium-doped silica glass, when pumping into the $^3F_4$ level. The white and grey circles denote the ions before and after the processes have taken place, respectively.

Figure 3. Log/log plot of the 800 nm luminescence vs. the 1800 nm luminescence, for pump powers ranging from 4 – 17 mW. Note: the errors associated with these measurements are not shown as they are smaller than the size of the individual data points presented in the plot.

Figure 4. Semi-log plots of the luminescence from the $^3H_4 \rightarrow {}^3H_6$ transition. The open circles represent the decay under indirect excitation at 1586 nm: The filled circles represent the decay under direct excitation at 786 nm. (a), (b) and (c) are the decay curves for samples with $Tm_2O_3$ concentrations of 35, 105 and 515 ppm mol, respectively. The least square fit is discussed in the text.